\documentclass[amssymb,amsmath,twocolumn,superscriptaddress]{revtex4-1}
\usepackage{makeidx}
\usepackage{amsmath}
\usepackage{amsfonts}
\usepackage{amssymb}
\usepackage[dvips]{graphicx}
\usepackage{epsfig}
\usepackage{geometry}
\usepackage{dcolumn}

\begin{document}

\title{Loschmidt echo in many-spin systems: a quest for intrinsic
decoherence and emergent irreversibility}
\author{Pablo R. Zangara}
\author{Horacio M. Pastawski} 
\address{Instituto de F\'{i}sica Enrique Gaviola (CONICET-UNC) and Facultad
de Matem\'{a}tica, Astronom\'{i}a, F\'{i}sica y Computaci\'{o}n (FaMAF), Universidad Nacional
de C\'{o}rdoba, 5000, C\'{o}rdoba, Argentina}



\begin{abstract}
If a magnetic polarization excess is locally injected in a crystal of
interacting spins in thermal equilibrium, this \textquotedblleft
excitation\textquotedblright\ would spread as consequence of spin-spin
interactions. Such an apparently irreversible process is known as spin
diffusion and it can lead the system back to \textquotedblleft
equilibrium\textquotedblright . Even so, a unitary quantum dynamics would
ensure a precise memory of the non-equilibrium initial condition. Then, if
at certain time, say $t/2$, an experimental protocol reverses the many-body
dynamics by changing the sign of the effective Hamiltonian, it would drive
the system back to the initial non-equilibrium state at time $t$. As a
matter of fact, the reversal is always perturbed by small experimental
imperfections and/or uncontrolled internal or environmental degrees of
freedom. This limits the amount of signal $M(t)$ recovered locally at time $%
t $. The degradation of $M(t)$ accounts for these perturbations, which can
also be seen as the sources of decoherence. This general idea defines the
Loschmidt echo (LE), which embodies the various time-reversal procedures
implemented in nuclear magnetic resonance. Here, we present an invitation to
the study of the LE following the pathway induced by the experiments. With
such a purpose, we provide a historical and conceptual overview that briefly
revisits selected phenomena that underlie the LE dynamics including chaos,
decoherence, localization and equilibration. This guiding thread ultimately
leads us to the discussion of decoherence and irreversibility as an emergent
phenomenon. In addition, we introduce the LE formalism by means of spin-spin
correlation functions in a manner suitable for presentation in a broad scope
physics journal. Last, but not least, we present new results that could
trigger new experiments and theoretical ideas. In particular, we propose to
transform an initially localized excitation into a more complex initial
state, enabling a dynamically prepared LE. This induces a global definition
of the LE in terms of the raw overlap between many-body wave functions. Our
results show that as the complexity of the prepared state increases, it
becomes more fragile towards small perturbations.
\end{abstract}

\maketitle

\section{Irreversibility and decoherence: a conceptual and historical
overview}

\subsection{On the emergent hierarchical structure of Nature}

Our knowledge about the Universe surrounding us is far from being
harmoniously unified. Instead, we have a large catalogue of almost independent
disciplines that constitute, with different degree of success, our
comprehension on how Nature works. Quantum Physics, Chemistry, Biology, and
their subfields belong to such a hierarchical catalogue that could end with
Psychology and Sociology. Each of these fields of knowledge provides partial
descriptions that correspond to different \textit{levels} of reality \cite%
{chibbaro2014Book}. The relation between the basic laws of a discipline in
terms of the ones ruling the previous level in the hierarchical tree
(usually called \textit{reduction}) is highly nontrivial as the\ passage
from one level to another may involve conceptual and mathematical
discontinuities or \textquotedblleft phase transitions\textquotedblright\ 
\cite{AndersonMORE,berry1994,primas1998}. Among these, life and
consciousness have proved to be the most elusive, with only a few hints as
to when and why they emerge \cite%
{koch2012consciousness,hofstadter2013strange}.

Within the realm of Physics, these transitions occur between different
domains characterized by specific energy, time and length scales. Here, we
have reasons to be more optimistic since we have well-developed experimental
and mathematical tools. To begin with, let us consider a textbook example
which is close to our central problem of irreversibility, namely the
derivation of equilibrium Thermodynamics from classical Statistical
Mechanics (SM). The former, one of the driving forces behind the Industrial
Revolution, describes the properties of matter in terms of pressure, volume,
temperature, among other macroscopic variables. The latter, mainly developed
by James C. Maxwell, Ludwig Boltzmann and Josiah W. Gibbs, provides a
probabilistic description of a system composed by $N$ \textquotedblleft
atomic\textquotedblright\ constituents. These elementary entities, and the
way in which they interact, are described by means of Classical Mechanics
(CM). In fact, Boltzmann \cite{BoltzmannBook,BrushBook} considered \textit{%
\textquotedblleft the world as a mechanical system of an enormously large
number of constituents, and of an immensely long period of
time\textquotedblright\ }$\left( t\right) $,i.e. what we currently
call the thermodynamic limit of $N\rightarrow \infty $ and then $%
t\rightarrow \infty $. Then, simple macroscopic relations (thermodynamic
equations of state) can be derived from SM by focusing on certain
microscopic observables and taking such a limit.

While the previous \textit{reduction} framework works in principle, and to a
large extent in practice, it fails when dealing with specific and highly
nontrivial physical situations: the critical points. Indeed, a critical
point indicates the onset of a phase transition, where the thermodynamic
variables, expressed as functions of certain control variable, become
non-analytic or even diverge. Here, the reduction paradigm has to be
replaced by the \textit{emergence} of qualitatively new phenomena. As stated
by Sir Michael Berry \cite{berry1994},

\textquotedblleft \textit{Thermodynamics is a continuum theory, so reduction
has to show that density fluctuations arising from interatomic forces have a
finite (and microscopic) range. This is true everywhere except at the
critical point, where there are fluctuations on all scales up to the sample
size. Thus, at criticality the continuum limit does not exist, corresponding
to a new state of matter. In terms of our general picture, the critical
state is a singularity of thermodynamics, at which its smooth reduction to
statistical mechanics breaks down; nevertheless, out of this singularity,
emerges a large class of new `critical phenomena', which can be understood
by a careful study of the large-}$N$\textit{\ asymptotes.\textquotedblright }

Our everyday experience corresponds to a macroscopically large classical
domain, where our falling glass shatters, our cup of coffee cools down, and
the ice cube left out of the refrigerator melts. All of these \textit{%
irreversible} phenomena are manifestations of the well known Second Law of
Thermodynamics, i.e. things evolve towards an unavoidable increase of
entropy. Indeed, the physical description of such a macroscopic domain
involves time-asymmetric equations of motion, such as hydrodynamic and
diffusive ones. As already stated, the elementary or microscopic
constituents of each macroscopic entity belong to a \textquotedblleft more
fundamental\textquotedblright\ level, whose physical description involves
time-symmetric equations of motion, such as Newtonian dynamics or, even
better, the Schr\"{o}dinger equation. In other words, while the microscopic
world is reversible, the macroscopic one is not. This paradoxical contrast
has fed a recurrent controversy since the end of the nineteenth century to
the present day. Our position on this issue relies on the emergence of
irreversibility from a microscopically reversible quantum dynamics provided
that the thermodynamic limit (TL) $N\rightarrow \infty $ is appropriately
considered.

This article constitutes an invitation to the study of irreversibility as en
emergent phenomenon following a pathway much entangled with a series of NMR
experiments led by Patricia Levstein and Horacio Pastawski at C\'{o}rdoba,
Argentina \cite{patricia98,usaj-physicaA,MolPhys}. In what follows, we
discuss and summarize the key concepts on this regard in a manner suitable
for a broad readership. Novel results and new insights of interest for
specialists are discussed in Sec. \ref{Sec_DPLE}.

\subsection{The \textit{irreversibility} \textit{paradox} in the classical
world}

In his attempt to reconcile the irreversible nature of the Second Law with
the reversible Newtonian laws of motion underlying SM, Boltzmann
considered the evolution of a gas composed by colliding particles which is
prepared out of equilibrium. Such a complex system was then described
according to a kinetic equation that irreversibly leads the system to
equilibrium. Here, the time-reversal symmetry is removed through the
assumption of \textquotedblleft molecular chaos\textquotedblright\ or 
\textit{stosszahl-ansatz}. This hypothesis implies that after each particle
collision (characterized by a typical time) the memory of the previous state
is lost. Boltzmann's approach finally led him to the celebrated $H$ theorem,
which is the first formal justification of the Second Law. At that time,
Joseph Loschmidt raised the objection that for every possible trajectory
that leads to equilibrium there will be another trajectory, equally
possible, that would lead to the initial out-of-equilibrium state.
Therefore, it would be possible to revert the velocities of every particle
to get again a low entropy state. As an answer, Boltzmann emphasized on the
extreme practical difficulty of achieving the time reversed evolution
proposed by Loschmidt, allegedly by saying \textquotedblleft \textit{it is
you who would invert the velocities!}\textquotedblright\ \cite{kuhn1978black}%
.

Boltzmann himself improved his theory transforming it into a probabilistic
description. More precisely, the separation between microscopic and
macroscopic scales, is exactly what enables us to predict the \textit{typical%
} evolution of a particular macroscopic system. In fact, this constitutes
the modern wisdom that explains the so called \textquotedblleft
irreversibility paradox\textquotedblright . As stated by Joel Lebowitz \cite%
{LebowitzRMP1999} (see also \cite{LebowitzPhysToday,LebowitzScholarpedia}),

\textquotedblleft \textit{... several interrelated ingredients which
together provide the sharp distinction between microscopic and macroscopic
variables required for the emergence of definite time-asymmetric behavior in
the evolution of the latter despite the total absence of such asymmetry in
the dynamics of individual atoms. They are: (a) the great disparity between
microscopic and macroscopic scales, (b) the fact that events are, as put by
Boltzmann, determined not only by differential equations, but also by
initial conditions, and (c) the use of probabilistic reasoning: it is not
every microscopic state of a macroscopic system that will evolve in
accordance with the second law, but only the `majority'\ of cases---a
majority which however becomes so overwhelming when the number of atoms in
the system becomes very large that irreversible behavior becomes a near
certainty.}\textquotedblright

The key point here is that, already in the classical Boltzmann's approach,
the notion of the \textit{emergent }irreversibility depends somehow on the
conditions under which the passage from the microscopic to the macroscopic
domain is performed. More precisely, the main \textit{qualitative} features
of this paradigm have only a slight dependence on the specific details of
the underlying microscopic dynamics. However, there exist microscopic
properties that do determine the \textit{quantitative} description of the
macroscopic evolution \cite{LebowitzRMP1999}. In other words, the derivation
of hydrodynamic and diffusive macroscopic equations ultimately depends on
the behavior of microscopic trajectories. These microscopic details can
include two mutually complementary properties:\ an extreme sensitivity
towards initial conditions, i.e. \textit{chaoticity}, and a tendency towards
a uniform distribution of the state in the available phase space, i.e. 
\textit{mixing}.

Let us explore the two ingredients just mentioned in the context of few-body
CM. According to CM, the state of a system, composed by $N$ particles in $d$
dimensions, is described as a point $X$ in a ($2dN$)-dimensional phase
space. If the system is conservative, the energy is the primary conserved
quantity, and the phase space is restricted to a hypersurface $\Omega $ of $%
2dN-1$ dimensions usually called \textit{energy shell}. In this scenario, 
\textit{chaos} implies that two trajectories in $\Omega $, starting from
points separated by an arbitrarily small distance $\delta _{0}$, will
separate exponentially as a function of time, $\delta (t)\sim \delta
_{0}\exp [\lambda t]$. Here, the typical inverse time $\lambda $ is a
Lyapunov characteristic exponent. The \textit{mixing} property implies that
the system evolves over time so that any given region contained in $\Omega $
eventually overlaps with any other given region in $\Omega $. This can be
thought as an \textquotedblleft intertwined\textquotedblright\ picture of
the phase space. None of these properties are satisfied in the case of fully
integrable systems, since their solutions are regular and non-dense orbits
in $\Omega $. But if integrability is completely broken, chaos and mixing
imply that the orbits become irregular and cover $\Omega $ densely. This
means that an actual trajectory $X(t)$ will be arbitrarily close to every
possible configuration within $\Omega $, provided that enough time has
elapsed. This last observation embodies the concept of \textit{ergodicity}:
an observable can be equivalently evaluated by averaging it for different
configurations in $\Omega $ or by its time-average along a single trajectory 
$X(t)$. Such a property is the cornerstone of -classical- SM since it sets
the equivalence between the Gibbs' description in terms of ensembles and
Boltzmann kinetic approach. Furthermore, we remark that this result does not
depend on the limit $N\rightarrow \infty $.

By the early 1950's, Enrico Fermi, John Pasta and Stanislaw Ulam (FPU) \cite%
{FPU} tried to study when and how the integrability breakdown could lead to
an ergodic behavior within a deterministic evolution. They considered a
string of harmonic oscillators perturbed by anharmonic forces in order to
verify that these nonlinearities can lead to energy equi-partition as a
manifestation of ergodicity. Even though Ulam himself stated
\textquotedblleft \textit{The motivation then was to observe the rates of
mixing and thermalization...}\textquotedblright\ \cite{fermi_book}, the
results were not those expected: \textquotedblleft
thermalization\textquotedblright\ dynamics did not show up at all. Instead,
the dynamics of the FPU problem, at least for the small number of particles
considered, evidenced remarkable recurrences much as those invoked by Henri
Poincar\'{e} in his famous recurrence theorem \cite%
{poincare1890probleme,Zermelo1,BrushBook}.

Nowadays, the striking FPU results are understood in terms of the
theory of chaos \cite{izraChaosReview}. More precisely, the microscopic
instabilities are systematically described by the theory of dynamical chaos
developed by Boris Chirikov \cite{chirikov1,chirikov2}. In fact, the onset
of dynamical chaos can be identified with the transition from non-ergodic to
ergodic behavior \cite{Zaslavsky}. A more general criterion for the onset of
ergodicity is given by the Kolmogorov-Arnold-Moser (KAM) theorem \cite%
{KAMscholarpedia}. It predicts that a weak nonlinear perturbation of an
integrable system destroys the constants of motion only locally in the
regions of resonances. In other regions of the phase space, islands of
quasi-periodic motion persist. Although a direct application of KAM theorem
to the FPU model suffers from technical difficulties, it constitutes the
main indication that one should not naively expect that weak nonlinear
perturbations ensure ergodicity. In addition, it is crucial to
remember that real physical systems are neither closed nor finite in a
strict sense. Already Boltzmann used this argument to conjure up the peril
that cyclic or recurrent motion posed on SM \cite{BoltzmannBook,BrushBook}:

\textit{\textquotedblleft In practice, however, the walls are continuously
undergoing perturbations, which will destroy the periodicity resulting from
the finite number of molecules\textquotedblright }.

Furthermore, the final way out should be found outside the realm of CM, in a
description already foreseen by Boltzmann, an emergent from a many-body
(quantum) description in the TL $N\rightarrow \infty $ \cite{BoltzmannBook,BrushBook}:

\textit{\textquotedblleft Since today it is popular to look forward to the
time when our view of Nature will have been completely changed, I will
mention the possibility that the fundamental equations for the motion of
individual molecules will turn out to be only approximate formulas which
give average values, resulting from the probability calculus from the
interactions of many independent moving entities forming the surrounding
medium--as for example metheorology laws are valid only on average values
obtained by long series of observations using the probability calculus.
These entities must be of course so numerous and must act so rapidly that
the correct average values are attained in millionths of a
second.\textquotedblright }{}

\subsection{A coherent quantum world}

Is it possible to formulate an straightforward extension of the previous
physical picture within Quantum Mechanics (QM)? Any closed and finite
quantum system involves a discrete energy spectrum and evolves
quasi-periodically in the Hilbert space, which becomes the quantum analogue
to the classical phase space. As in CM, a closed and finite quantum system
is intrinsically reversible due to the unitarity of the evolution operator.
However, there exists a crucial difference: while integrability, ergodicity
and chaos are well-established concepts within CM, their extensions in QM
are much less clear.

The notion of \textit{integrability} in the QM literature may refer to
different criteria. Contrary to CM, the existence of \textquotedblleft $N$
independent (local) conserved mutually commuting linearly independent
operators\textquotedblright\ does not necessarily imply that the system is
\textquotedblleft exactly solvable\textquotedblright\ in the quantum domain\ 
\cite{Mossel2011,Eisert2011}. In addition, the concept of \textit{ergodicity}
in QM has also generated intense debate, even after the recent rediscovery
of the John von Neumann's Quantum Ergodic Theorem \cite%
{vonNeumannQET,LebowitzQET}. Last, but not least, the classical definition
of chaos as the sensitivity to initial conditions does not apply to a wave
equation as Schr\"{o}dinger's. Indeed, we will discuss below that the
quantum signature of dynamical chaos had to be found as an instability of a
quantum evolution towards perturbations in the Hamiltonian \cite{jalpa}.

\subsubsection{Decoherence and the theory of open quantum systems}

A phenomenological description of irreversible dynamics within QM can be
performed by postulating that the system of interest is in contact with an
environment. This implies the removal of the assumption that the system is
\textquotedblleft closed\textquotedblright . As stated by G\"{o}ran Lindblad 
\cite{Lindblad1976},

\textquotedblleft \textit{It seems that the only possibility of introducing
an irreversible behavior in a finite system is to avoid the unitary time
development altogether by considering non-Hamiltonian systems. One way of
doing this is by postulating an interaction of the considered systems S with
an external system R like a heat bath or a measuring instrument.}%
\textquotedblright

The first consequence of the coupling to infinitely many environmental
degrees of freedom seems to be the destruction of quantum weirdness within
the system. More precisely, this means the attenuation of the interferences
evidenced in specific observables and, ultimately, the appearance of
classicality \cite{zurek}. Since this degradation is originated in the loss
of phase-coherence between the components of specific quantum
superpositions, such a process is called 
\textit{decoherence}. This automatically implies that the open system
dynamics is irreversible.

A theoretical framework, developed in the 1960's by Leo Kadanoff and Gordon
Baym \cite{Kadanoff_book} and by Leonid Keldysh \cite{keldysh}, describes
the non-equilibrium SM and it intrinsically deals with open-system dynamics.
Such a framework uses the tools of Quantum Field Theory and it ultimately
involves the TL. Precisely, this limit may shadow the nature of the
approximations involved. A more \textquotedblleft
controlled\textquotedblright\ formalism to describe open-system dynamics had
to wait until 1976, when Lindblad's work \cite{Lindblad1976} was
independently and simultaneously complemented by Vittorio Gorini, Andrzej
Kossakowski, and George Sudarshan \cite{GKS1976}. They provide the
mathematical structure of the quantum master equations (QME), which in turn
can be understood as generalized Liouville-von Neumann differential
equations for the reduced density matrix. In this approach, the system under
study is finite and specific assumptions are made to describe its
environment. These physical hypothesis, which are basically known as the
Markovian approximation, were already known and discussed in the 1950's by
Felix Bloch \cite{Bloch1953} and Ugo Fano \cite{Fano1957}. As summarized by
Karl Blum \cite{blum-1996},

\textquotedblleft \textit{It is assumed that R has so many degrees of
freedom that the effects of the interaction with S dissipate away quickly
and will not react back onto S to any significant extent so that R remains
described by a thermal equilibrium distribution at constant temperature,
irrespective of the amount of energy and polarization diffusing into it from
the system S. In other words, it is assumed that the reaction of S on R is
neglected and the correlations between S and R, induced by the interaction,
are neglected.}\textquotedblright

The QME approach has proved to be operationally successful in describing the
dynamics of open systems, as in the case of nuclear magnetic resonance (NMR) 
\cite{munowitz1988coherence,Ernst-Book} and quantum optics\textbf{\ }\cite%
{zoller_book,Petruccione_book}.\ The more prominent model within the theory
of open quantum systems is the two-level system in the presence of a
structured environment \cite{LeggettRMP}. In particular, the
Kadanoff-Baym-Keldysh approach was recently applied to such a model with the
specific purpose of describing \textit{quantum dynamical phase transitions} 
\cite{hmp-physB}. This novel phenomenon, firstly observed in NMR \cite%
{QDPT_exp}, corresponds to a functional change or non-analyticities in the
dynamics of specific observables. More recently, these transitions were also
described by means of the QME approach \cite{CiracPRA2013}. In any case, the
environment is required to be already in the TL, a requisite automatically
satisfied in the experimental realm.

\subsubsection{Closed quantum systems: to equilibrate or to localize? \label{Sec_eqloc}}

As in the original Boltzmann's gas of colliding molecules, let us consider
now a closed quantum system composed by a large number $N$ of interacting
particles, e.g. fermions or spins in a lattice. We assume that the initial
state of the system ($t=0$) is given by $\hat{\rho}_{0}$, which does not
commute with the time-independent Hamiltonian $\hat{H}$. In other words, $%
\hat{\rho}_{0}$ is a non-equilibrium state. Its time evolution is given by 
\begin{equation*}
\hat{\rho}_{t}=\exp \left[ -\frac{\mathrm{i}}{\hbar }t\hat{H}\right] \hat{%
\rho}_{0}\exp \left[ \frac{\mathrm{i}}{\hbar }t\hat{H}\right] ,
\end{equation*}%
which, being dependent on time through quasiperiodic functions, is nearly
recurrent and, strictly speaking, it cannot describe the approach to any
\textquotedblleft equilibrium\textquotedblright\ state. This is merely the
result of an unitary dynamics.

Given a specific observable, say $\hat{O}$, the time-evolution of its
expectation value is: 
\begin{equation*}
\langle \hat{O}(t)\rangle =\mathrm{tr}\left( \hat{\rho}_{t}\hat{O}\right) .
\end{equation*}%
Surprisingly, provided that $\hat{O}$ fulfils certain conditions, \textit{%
equilibration} can occur for $\langle \hat{O}(t)\rangle $. Here,
equilibration means that after some transient behavior, $\langle \hat{O}%
(t)\rangle $ reaches some \textquotedblleft stationary\textquotedblright\
value and remains \textit{close} to it for \textit{most} of the time \cite%
{vonNeumannQET,LebowitzQET}. There exists several theoretical and
experimental questions around such idea \cite{Eisert2014,Eisert2016}. For
instance, one may naturally ask on the conditions required for $\hat{O}$, $%
\hat{H}$, and $\hat{\rho}_{0}$ in order to observe the equilibration of $%
\langle \hat{O}(t)\rangle $. Indeed, it seems that a key requirement for
equilibration is the \textit{locality} of $\hat{O}$. Loosely speaking, this
means that $\hat{O}$ involves only a small number of sites or particles. The
idea of locality provides for a natural argument in favor of an apparently
irreversible behavior of $\langle \hat{O}(t)\rangle $. Indeed, if we
\textquotedblleft observe\textquotedblright\ a local subsystem that involves
only a small fraction of the degrees of freedom of the entire system, the
coupling to the rest of the system mimics a coupling to an environment.
Therefore, as in the above discussion on open system dynamics, the
relaxation of $\langle \hat{O}(t)\rangle $ occurs due to the presence of a
large environment \textquotedblleft observing\textquotedblright\ the
subsystem. In such a case, the complete system is said to act as its own
environment.

Given an interacting many-body system, is it natural to expect that local
observables do equilibrate? The answer is no, since there exist striking
phenomena in which the equilibration mechanisms are completely inhibited. In
fact, a paradigmatic example can be drawn from a very active area in
physics: the \textit{quantum localization}. The concept was
originally developed by Philip W. Anderson \cite%
{Anderson1958}, who described the absence of diffusion of spin excitations
in a disordered system. This phenomenon was observed by George Feher in
electron-nuclei double resonance experiments performed in doped
semiconductors \cite{AndersonRMP-1978}. While the problem had indeed a
many-body nature, Anderson succeed in simplifying it as a system of
non-interacting electrons in a $d$-dimensional disordered lattice. He realized
that its dynamics described by Bloch states can change dramatically when a
perturbing disordered potential exceeds a critical value. This is known as
the extended-to-localized transition or Anderson's localization (AL) \cite%
{AndersonRMP-1978}. Indeed, if the disorder is small, single-particle states
are essentially described in terms of the scattering of Bloch states with
some finite lifetime. Precisely, this physical picture was employed by
Robert Laughlin to re-examine quantum transport in a random potential as a problem of Quantum Chaos \cite{Laughlin1987}. He proved that the Lyapunov characteristic exponent of the classical electron motion in such a potential can be identified with the collision rate $1/\tau$ appearing in Ohm's law. This constitutes a conceptual link between chaos and diffusive transport. Then, for small disorder, one can generally say that a local excitation is described as a wave packet that ultimately diffuses throughout the system. But AL involves a paradigm shift: when the disorder is large
enough, the eigenstates of the system become localized in the real space. Thus, in contrast to the diffusive picture, localization implies that the excitation remains close to its initial location and transport phenomena is no longer possible \cite{MacKinnon93,Pastawski1985}. In some sense, the same impurities that produce
chaos and Lyapunov exponents, which make possible diffusion, dissipative
transport and equilibration \cite{Laughlin1987}, end up conspiring against
them. For dimension $d\leq 2$, even a weak disorder ensures the onset of AL. 

A step beyond the standard AL problem corresponds to the original problem of
localization in interacting systems. As a matter of fact, adding
interactions increases the effective dimensionality of the problem, and hence significantly magnifies its complexity. A clean
(ordered) system can already evidence a transition into an insulating phase
if the interactions are strong enough. This is the case of the Mott-Hubbard transition, whose most paradigmatic case occurs
in a crystal with one electron per atomic orbital. John Hubbard showed by means of a Hartree-Fock calculation
that for strong local electron-electron repulsion, an otherwise half-filled band of electronic
states would split into an occupied band
and an unoccupied one \cite{Hubbard1963}. This drives the system
from a metallic phase to a insulator one  \cite{MottRMP,madelungBOOK}, a situation that would persist at fillings
where the strong electron-electron Coulomb repulsion forbids the identification of single-particle bands. If the system is disordered, the
naive expectation would be that interactions prevent the existence of AL
essentially for two reasons. First, collisional dephasing would destroy the
specific interferences needed to localize. Second, including interactions
automatically implies an exponential increase in the fraction of the Hilbert
space required to describe the excitation dynamics. This, in turn,
constitutes a more favorable scenario for diffusion, consistent with the
fact that a large $d$ prevents localization. Nevertheless, a different kind
of transition does occur between extended and localized many-body phases,
which is called Many-Body Localization (MBL) \cite%
{altshuler2006,Altshuler2010}. The MBL is a dynamical transition that
results when both perturbations to Bloch states, i.e. interactions and
Anderson disorder, are present. As in the AL case, an interacting many-body
system is said to be localized if the diffusion or transport of excitations
becomes frozen, and therefore a memory of the initial conditions is
preserved in local observables for long times \cite{Huse2014}. However,
there is a hierarchical difference between AL and MBL: while the former
deals with the eigenstates in a single particle Hilbert space, the latter
relies on the properties of the many-body eigenstates in the much bigger
Fock space.

The study of localization phenomenon, and particularly the MBL, is extremely
attractive from fundamental grounds. Its importance traces back to the FPU
problem and the (classical) theoretical framework developed to understand
it. As in the attempt by Fermi and collaborators, the current aim is to
study simple quantum models that could go parametrically from an ergodic to
a non-ergodic quantum dynamics. Moreover, a fundamental question is whether
such a transition occurs as a smooth crossover or it has a critical value, a
sort of generalized quantum KAM threshold \cite{polkovnikovRMP}. As a matter
of fact, the MBL transition is the promising candidate in the quantum realm.
If the many-body states are extended, then one may expect that the system is
ergodic enough to behave as its own environment, and, as stated before,
equilibration is enabled. Quite on the contrary, if the many-body states are
localized, any initial excitation would remain out-of-equilibrium. In this
case, equilibration is precluded. Therefore the MBL would evidence the
sought threshold between ergodic and non-ergodic behaviors.

During the last years, the literature dealing with the dynamics of
equilibration and thermalization has grown overwhelmingly (see Refs. \cite%
{Eisert2016} and \cite{Huse2014}). On the experimental side, an extreme
degree of isolation and control has been achieved, such as in the case of
ultracold quantum gases \cite{Bloch_review2012}, trapped ions \cite%
{Blatt_review2012} and also NMR \cite{gonzaloRMP}. On the theoretical side,
the use of the Hubbard model \cite{Hubbard1963} has proven useful to address
the dynamics of strongly-interacting many-body systems \cite%
{Tomadin2007,Buonsante2008,Shepelyansky2016}. In addition, for spin systems, numerical studies have provided evidence 
indicating that the competence between disorder and interactions may lead to highly non-trivial dynamical phase diagrams \cite{Zangara2013PRB,Zangara2015PIP}.

\subsection{Echoes from the future: NMR comes back}

We are now ready to focus on our major question. Namely, to what extent
would equilibration, as introduced above, could turn the excitation
spreading into an irreversible phenomenon? In order to tackle such a
question, we consider specifically a system composed by $N$ quantum spins
which is perturbed from a high temperature equilibrium by injecting a local
polarization excess. In addition, we assume that the system can indeed
equilibrate, i.e. dynamics can lead to a homogeneous distribution of the
polarization. Such an equilibration of the polarization is a consequence of
a mechanism usually called \textit{spin diffusion} \cite%
{BlumeHubbard1970,forsterbook}. However, even though a particular observable
seems to have reached \textquotedblleft equilibrium\textquotedblright , a
unitary quantum dynamics would ensure a precise memory of the
non-equilibrium initial condition. In other words, the initial state is
completely encoded into correlations (eventually non-local) present in the
evolved state.

If some experimental protocol could manage to achieve the inverse evolution
operator, i.e. to \textit{reverse} the many-body dynamics, then it would
drive the system back to the initial non-equilibrium state \cite%
{Hahn_atomicMemory}. This \textquotedblleft echo idea\textquotedblright\ has
remained at the heart of NMR. In fact, the first NMR time-reversal
experiments were performed by Erwin Hahn in the 1950's \cite{Hahn1950}.
There, the total polarization, that sums up contributions from individual
spins, when lying perpendicular to the external magnetic field, will ideally
precess indefinitely. However, each independent spin contributing to the
polarization precesses at different velocity\ due to the local
inhomogeneities of the magnetic field. This dephasing produces the decay of
the observed polarization. Hahn's procedure involves the inversion of the
sign of the Zeeman energy in order to reverse these precessions. If such an
inversion is performed at time $t/2$, it will produce the refocusing of the
total polarization at time $t$, which is the well known \textquotedblleft
spin echo\textquotedblright . Nevertheless, the\ sign of the energy scale
associated to the spin-spin interactions is not inverted and, accordingly,
the echo signal $M(t)$ becomes progressively degraded with $t.$ This decay
has a characteristic time scale called $T_{2},$ which, in a crystal,
characterizes these many-body interactions.

By the early 1970's, Horst Kessemeier, Won-Kyu Rhim, Alex Pines, and John
Waugh implemented the reversal of the dynamics induced by the spin-spin
dipolar interaction which was missed by the Hahn's procedure \cite%
{Kessemeier1971,Rhim1971}. This results in the \textquotedblleft magic
echo\textquotedblright\ signal, which indicates the recovery of a global
polarization state. Two decades later, Richard Ernst and collaborators
introduced the \textquotedblleft polarization echo\textquotedblright\ \cite%
{Ernst1992}. There, the polarization along the external field is injected 
\textit{locally} at some labeled spins. As the global polarization is a
conserved quantity in this experiment, the local polarization diffuses away
for a time $t/2$ due to the spin-spin dipolar interaction. Since it
originates in many-spin interactions, this spreading occurs in time
scale $T_{2}$. By the implementation of a time-reversal procedure involving
a change in the sign of the spin-spin energy at time $t/2$, some
polarization $M(t)$ is recovered at time $t$ in the same initial spots where
it was injected.

While the success of these time reversal echoes unambiguously evidenced the
deterministic and quantum nature of spin-dynamics in NMR, it is unavoidable
that the reversal results degraded by uncontrolled internal or environmental
degrees of freedom, or by imperfections in the pulse sequences. In terms of
the theory of open quantum systems, these perturbations would constitute
sources of decoherence. Quite often, their precise assessment could only be
evidenced through the time reversal experiment \cite%
{patricia98,HMP_RevMex1998}. In addition, it was noticed that, since in the
\textquotedblleft polarization echo\textquotedblright\ experiments the
total polarization is a conserved quantity, an imperfectly recovered $M(t)$\ would
be roughly proportional to inverse of the number of spins among which the
polarization is distributed. This gives a clear entropic meaning for $-\ln
[M(t)]$\cite{MolPhys}. Most importantly, the degradation of $M(t)$ in such\
experiments, where the many-spin interactions are globally and quite
perfectly time-reversed, occurs in a new time scale $T_{3}$. Remarkably, the
experiments indicate that $T_{3}$ seems to be much shorter than any naive
estimation of the characteristic time scale of the perturbations, say $\tau
_{\Sigma }$. Thus, the question is whether the complexity inherent to a
large number of correlated spins would enhance the fragility of the
procedure under perturbations \cite{patricia98}.

\subsection{The Loschmidt echo: irreversibility as an emergent phenomenon \label{Sec_LEdiscussion}}

The time-reversal procedures just described can be thought as a quantum
realization of the gedanken experiment suggested by Loschmidt. As mentioned
above, he proposed the reversal of the particles' velocities as a mechanism
for undoing the increase of entropy. Actually, the aforementioned NMR
time-reversal experiments are specific procedures for backtrace the time
evolution \cite{Waugh1972}. This led Patricia Levstein and Horacio Pastawski
to define the \textquotedblleft \textit{Loschmidt echo}\textquotedblright\
(LE) as an idealization which embodies the various, eventually imperfect,
time-reversal procedures implemented in NMR \cite{scholarpedia}.

A next generation of experiments in organic crystals \cite%
{patricia98,usaj-physicaA,MolPhys}, suggested that the experimental time
scale for irreversibility $T_{3}$ never exceeds more than a few times $%
T_{2}, $ and indeed, it remains tied to it \cite{usaj-physicaA,MolPhys}.
Thus, it seems that one is left with a sort of unbeatable limit, 
\begin{equation*}
T_{2}\lesssim T_{3}\ll \tau _{\Sigma }.
\end{equation*}%
The most immediate conclusion would be that there is still an uncontrolled
source of decoherence not described by $\tau _{\Sigma }$, an explanation
that both theoreticians and experimentalists would subscribe to \cite%
{zurekpolonica}. However, the experiments hinted that this \textquotedblleft
perturbation independent decay\textquotedblright\ (PID)\ has a much deeper
origin, with $T_{3}$ acting as a sort of inverse Lyapunov exponent \cite%
{usaj-physicaA}. Since this is in fact an intrinsic property of the system
(in the absence of external perturbations), the PID can also be understood
as an \textit{intrinsic decoherence} rate. We should stress that until now,
this observation holds for LE experiments where polarization is a conserved
magnitude. Those of the magic echo type fail to fulfill this requirement 
\cite{Lichi2015}.

\begin{figure*}
\centering
\includegraphics[width=0.6\textwidth]{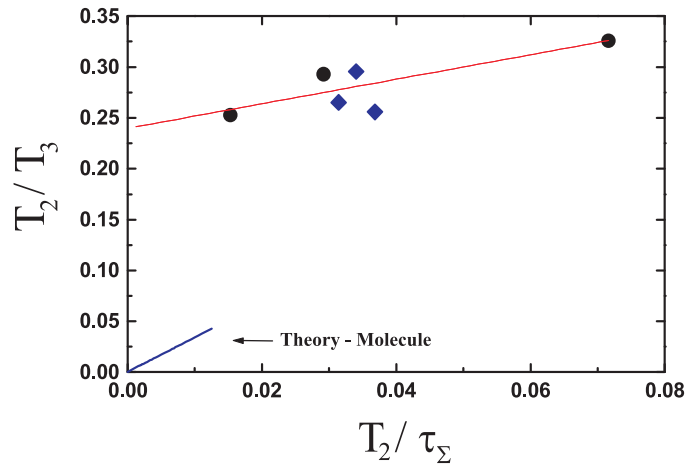} 
\caption{The LE decay rate $1/T_{3}$ as a
function of the perturbation's characteristic rate $1/\protect\tau _{\Sigma
} $. Both quantities are rescaled by the $T_{2}$ time scale, which
corresponds to the reversible or controlled Hamiltonian. Thus, the vertical
axis corresponds to the irreversibility time-scale, while the horizontal one
is associated with the strength of the perturbation. In the limit of
vanishing perturbation, the irreversibility time-scale saturates at a
fraction $\sim \frac{1}{4}$ of the reversible time-scale. This drastically
differs from what expected for a single molecule: a perfectly reversible
dynamics as the perturbation goes to zero. Adapted from \protect\cite%
{UsajTesis}.}
\label{PIDexperimental}
\end{figure*}

When the decay rate $1/T_{3}$ (of a LE with polarization conserved) is
plotted against the intensity of the most relevant non-inverted term, namely
the residual non-secular interaction, it saturates to a finite value: $%
T_{3}\sim 4T_{2}$. This is explicitly shown in Fig. \ref{PIDexperimental}.
This plot resembles a standard resistivity vs. temperature plot in an impure
metal, where the finite resistivity offset at the zero temperature limit is
determined by the impurity scattering \cite{DoniachBook}, i.e. by chaos \cite%
{Laughlin1987}. Analogously, $T_{3}$ in the zero perturbation limit, keeps
tied to the time scale that characterizes the reversible many-body
interaction. This observation led to postulate the \textit{Central
Hypothesis of Irreversibility}: in an infinite many-spin system far away
from its ground state, the complex dynamics amplifies the action of any
small non-inverted interaction to the degree that such complex (reversible)
dynamics provides for the dominant time-scale. Thus, reversible interactions
responsible for spin diffusion\ turn out to be the determinant contribution
to the irreversibility rate.


The previous hypothesis triggered the theoretical study of simpler problems
in which the LE can be evaluated systematically. In particular, the role of
a \textquotedblleft testing bench\textquotedblright\ was played by single
particle systems whose classical counterpart is chaotic. There, the LE is
defined as \cite{jalpa}%
\begin{equation}
M(t)=\left\vert \left\langle \psi \right\vert \exp [\frac{\mathrm{i}}{\hbar }%
(\hat{H}_{0}+\hat{\Sigma})t]\exp [-\frac{\mathrm{i}}{\hbar }\hat{H}%
_{0}t]\left\vert \psi \right\rangle \right\vert ^{2},  \label{eco_jalpa}
\end{equation}%
i.e. the square of the overlap between two one-body wave functions, one of
them being evolved by an unperturbed Hamiltonian $\hat{H}_{0}$ and the other
by a perturbed one $\hat{H}_{0}+\hat{\Sigma}.$ A pioneering prediction by
Asher Peres \cite{peres1984} pointed that, if the unperturbed
evolution is classically chaotic, the long time limit $M(\infty )$ would
yield an evenly spread excitation (not expected for integrable systems).
Intermediate times probed a more subtle behavior. Semiclassical calculations
showed that at a finite energy $\varepsilon $ and small perturbation
strengths (but exceeding the spectral discreteness), the LE decay rate $%
1/T_{3}\ $equals $\Gamma (\Sigma ,\varepsilon )$, i.e.\ the broadening
estimated from a Fermi golden rule (FGR) calculation \cite%
{jalpa,Jacquod-Beenakker-2001}\textbf{. }This lifetime is proportional to
the square of the perturbation strength. Exponential decays were soon found
for different observables and methods \cite{Prosen2002PRE,Tomsovic2002}.
However, the real surprise was the existence of a regime in which the decay
rate of the LE corresponds to the classical Lyapunov exponent \cite%
{jalpa,Cookprb2004}. More specifically 
\begin{equation*}
1/T_{3}=\min [\Gamma /\hbar ,\lambda ].
\end{equation*}%
The LE Lyapunov regime is a particular PID that holds for a semiclassical
initial state built from a dense spectrum with a perturbation above certain
critical threshold $\Sigma _{c}$. Notably, $\Sigma _{c}$ falls inversely
proportional to the energy $\varepsilon $ of the state. Since $\varepsilon
\rightarrow \infty $ is equivalent to $\hbar \rightarrow 0$, we may say that 
$\Sigma _{c}$ vanishes in the classical limit, providing the elusive
quantum-classical limit \cite{PazZurek2003}.

One should point out that the LE Lyapunov regime may not be fully equivalent
with the more natural Lyapunov growth of correlation functions at short
times which is, by now, much used to address the so called \textquotedblleft
information paradox\textquotedblright\ of black holes in the context of
AdS-CFT \cite%
{Susskind2008,Stanford_butterfly,Susskind2015,maldacena2015bound,Kitaev_charla,swingle2016}%
. In fact, the search for diverging correlations was inspired by the
expected growth of the uncertainties of an electron wave packet propagating
in a dirty metal \cite{larkin1969}. Such an evolution of the correlation
functions holds up to the Ehrenfest or \textquotedblleft
scrambling\textquotedblright\ time, when different portions of a spread wave
packet are scattered by different impurities. At this point we should
remember that the LE Lyapunov regime holds until the much longer Thouless
time, when the already scrambled excitation had fully spread through the
whole system \cite{Cookprb2004,CookTesis}.

The discovery of the PID or Lyapunov regimes in specific LE evaluations was
a very big leap that triggered the interest of the Quantum Chaos and Quantum
Information communities \cite{prosen,Jacquod}. Indeed, the mere existence of
a PID regime would pose a great challenge on the controllability of quantum
devices as it evidences an intrinsic fragility of quantum dynamics towards
minuscule perturbations. The sensitivity to perturbations or fragility of
quantum systems \cite{Jacquod2002,Zurek2002,Bendersky2013} has become a
major problem that transversely affects several fields, e.g. possible
chaoticity of quantum computers \cite{Shepelyansky2000,Flambaum2000a}, NMR
quantum information processing \cite%
{Suter2004,Boutis2012,Cappellaro2013,Claudia2014}, quantum criticality \cite%
{Zanardi2006,Silva2008} and, more recently, quantum control theory \cite%
{Calarco2014}. Nevertheless, the ultimate numerical or analytical proof that
would support the above hypothesis, and thus explain the experimental
observations, is still lacking. The reason for such a theoretical bottleneck
relies on the inherent difficulty in addressing many-body dynamics.
Precisely, the association of many-body complexity with a form of chaos \cite%
{Flambaum2000,Flambaum2001d}, could provide a rational for the
experimentally observed PID.

An unavoidable starting point for the analysis of PID is the identification
of the conditions for the \textquotedblleft perturbation \textit{dependent}
decay\textquotedblright\ that manifests at the relatively short times when
exponential decay described by FGR is manifested \cite{Zangara2012}. As
mentioned above, in single-particle systems with a semiclassical excitation,
the FGR holds for weak perturbations up to threshold given by the classical
Lyapunov exponent that indicates the onset of the PID. So, could a many-body
system have a vanishing threshold for the PID? If that is the case, it could
show up as a quantum dynamical phase transition in the TL ($N\rightarrow
\infty $). As a matter of fact, a standard experiment involves a crystalline
sample with an infinitely large number of spins. In other words, experiments
are \textquotedblleft already in the TL\textquotedblright . In contrast, any
numerical approach to assess many-spin dynamics can only cope with a
strictly finite $N$. This is the key point where the original discussion on
an emergent irreversibility comes back into scene: an appropriate finite
size scaling is required in order to grasp the emergent mechanism that rules
irreversibility in the TL. Indeed, one should increase progressively $N$\
going from small systems to larger ones with a controlled perturbation.
There, the emergent behavior would follow only in a precise order of the
limits: first $N\rightarrow \infty $\ and then $\tau _{\Sigma }\rightarrow
\infty $.

Reference \cite{Zangara2015PRA} constitutes a first attempt to pursue such
an ambitious program\textbf{. }The numerical detection of the PID was not
achieved and, in fact, it may stay beyond the state-of-the-art numerical
techniques \cite{Zangara2016}. However, an effective FGR regime was reported
being consistent with an emergent picture of irreversibility. Keeping in
mind the crucial order of the limits stated above, in an infinitely large
system, an infinitesimal perturbation is associated to a finite $\tau
_{\Sigma }$. If a \textit{unitary evolution} keeps the polarization
equilibrated during a time $t>$ $\tau _{\Sigma }$, such an equilibration
becomes \textit{irreversible}, and its time-reversal would be completely
ineffectual.

In what follows, inspired by the LE experiments in NMR, we provide a
framework for theoretical evaluations of LE of the polarization type. We do
not consider here LE of the magic echo type, as in this case neither the
numerical resolution of small systems \cite{Fine2014} nor the actual
experiments on scaled Hamiltonians \cite{Lichi2015} seem to evidence a PID
regime. We start by summarizing the LE formulation in spin systems as a
local autocorrelation function. Moreover, as introduced in Ref. \cite%
{Zangara2016}, we discuss the local correlation in terms of two global ones.
One of these global correlations is defined as the \textit{many-body} LE and
resembles the standard one-body LE definition embodied by Eq. (\ref%
{eco_jalpa}). The other is an initially fast growing multi-spin cross
correlation that later decays. This formalization leads us to address a
fundamental question underlying the LE literature \cite%
{Jalabert2016PhilTrans}:\ what is the relation between the LE as defined in
one-body systems, i.e. the overlap of wave functions, and the LE as defined
in many-body systems, i.e. a spin correlation function? Precisely, we
propose and discuss a dynamical preparation protocol to transform the local
LE into a global one. Such a dynamically prepared LE (DPLE) can, in turn,
lead to a new series of NMR experiments to systematically address the
fragility of strongly correlated many-body systems in the presence of small
perturbations.

\section{The Loschmidt echo formulation}

\subsection{The local autocorrelation function}

Following the early LE experiments \cite{patricia98,usaj-physicaA,MolPhys},
let us discuss and formalize here the spin system and the time reversal
protocol. We consider $N$ spins $1/2$ whose initial condition is described
by an infinite temperature state, i.e. a completely depolarized mixture,
plus a locally injected polarization,%
\begin{equation}
\hat{\rho}_{0}=\frac{1}{2^{N}}(\mathbf{\hat{I}}+2\hat{S}_{1}^{z}).
\label{rho_inicial}
\end{equation}%
Here, the spin $1$ is polarized while the others are not, i.e. $tr[\hat{S}%
_{i}^{z}\hat{\rho}_{0}]=\frac{1}{2}\delta _{i,1}$.

The polarization, which is initially placed in a single spin, diffuses all
around due to the spin-spin interactions. More precisely, a many-spin
Hamiltonian $\hat{H}_{0}$ rules such a \textit{forward} evolution of the
system up to a certain time $t_{R}$. At that moment, an inversion of the
sign of $\hat{H}_{0}$ is performed, leading to a symmetric \textit{backward }%
evolution. Typically, $\hat{H}_{0}$ stands for a truncated dipolar
Hamiltonian. Nevertheless, there are unavoidable perturbations, denoted by $%
\hat{\Sigma}$, that could arise from the incomplete control of the spin
Hamiltonian, acting on both periods. The procedure ends up with a local
measurement in the same spin that was originally polarized. See Fig. \ref%
{Fig_esquemaLE}. The evolution operators for each $t_{R}$-periods are $\hat{U%
}_{+}^{{}}(t_{R})=\exp [-\frac{\mathrm{i}}{\hbar }(\hat{H}_{0}+\hat{\Sigma}%
)t_{R}]$ and $\hat{U}_{-}^{{}}(t_{R})=\exp [-\frac{\mathrm{i}}{\hbar }(-\hat{%
H}_{0}+\hat{\Sigma})t_{R}]$ respectively. Thus, it is natural to define the
LE\ operator as:%
\begin{equation}
\hat{U}_{LE}^{{}}(2t_{R})=\hat{U}_{-}^{{}}(t_{R})\hat{U}_{+}^{{}}(t_{R}),
\label{Ule}
\end{equation}%
which produces an imperfect refocusing at time $2t_{R}$. The local
measurement of the polarization, performed at site $1$, defines the local LE:%
\begin{equation}
M_{1,1}(t)=2tr[\hat{S}_{1}^{z}\hat{U}_{LE}^{{}}(t)\hat{\rho}_{0}\hat{U}%
_{LE}^{\dag }(t)]=2tr[\hat{S}_{1}^{z}\hat{\rho}_{t}].
\label{autocorrelacion1}
\end{equation}%
Here, we choose as free variable $t=2t_{R}$, the total elapsed time in the
presence of the perturbation. The time dependence of $\hat{\rho}_{t}$ in the
Schr\"{o}dinger picture is,%
\begin{equation}
\hat{\rho}_{t}=\hat{U}_{LE}^{{}}(t)\hat{\rho}_{0}\hat{U}_{LE}^{\dag }(t).
\label{autocorrelacion11}
\end{equation}


\begin{figure*}[!ht]
\centering
\includegraphics[width=0.85\textwidth]{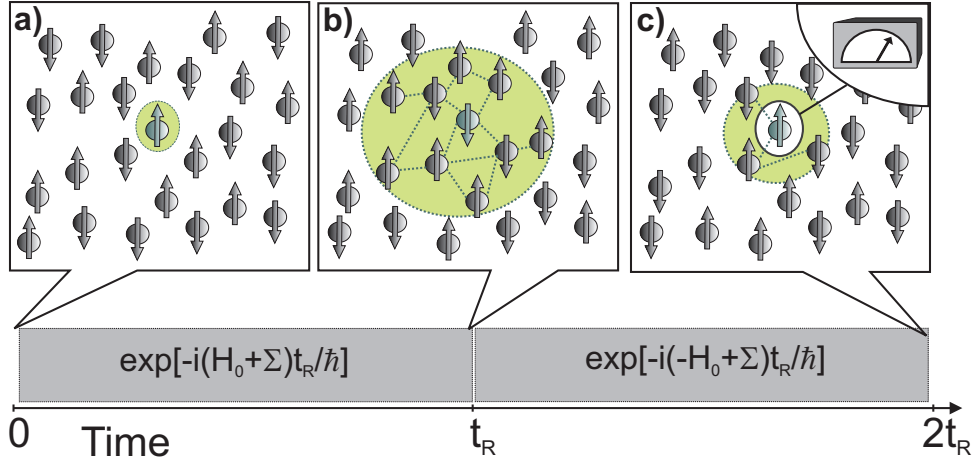} 
\caption{The pictorial scheme of the
dynamics involved in $M_{1,1}(t)$. (a) A local excitation in injected in a
high temperature spin system. The corresponding state is given by Eq. (%
\protect\ref{rho_inicial}). The system evolves ruled by the Hamiltonian $%
\hat{H}_{0}+\hat{\Sigma}$. Spin diffusion leads to the spreading of the
excitation until a time $t=t_{R}$ (b). At that time, the sign of $\hat{H}%
_{0} $ is inverted. A backward evolution takes place ruled by $-\hat{H}_{0}+%
\hat{\Sigma}$. At time $t=2t_{R}$ (c), at the same initial spot, a local
measurement is performed.}
\label{Fig_esquemaLE}
\end{figure*}


Using Eq. (\ref{rho_inicial}), and after some algebraic manipulation, the LE
can be explicitly written as a correlation function:%
\begin{eqnarray}
M_{1,1}(t)&=&\frac{1}{2^{N-2}}tr[\hat{U}_{LE}^{\dag }(t)\hat{S}_{1}^{z}(0)%
\hat{U}_{LE}^{{}}(t)\hat{S}_{1}^{z}(0)]  \notag \\
&=&\frac{tr[\hat{S}_{1}^{z}(t)\hat{S}_{1}^{z}(0)]}{tr[\hat{S}_{1}^{z}(0)\hat{%
S}_{1}^{z}(0)]}.  \label{autocorrelacion2}
\end{eqnarray}%
Here, the time dependence is written according to the Heisenberg picture,%
\begin{equation}
\hat{S}_{1}^{z}(t)=\hat{U}_{LE}^{\dag }(t)\hat{S}_{1}^{z}(0)\hat{U}%
_{LE}^{{}}(t).  \label{autocorrelacion21}
\end{equation}%
Notice that Eq. (\ref{autocorrelacion2}) is an explicit correlation function
at the same site but different times, i.e. an \textit{auto}correlation. This
correlation has been recently employed to address many-body localization in
spin systems \cite{Zangara2013PRB,Zangara2015PIP}.

If we use the identity $\hat{S}_{1}^{z}=\hat{S}_{1}^{+}\hat{S}_{1}^{-}-\frac{%
1}{2}\mathbf{\hat{I}}$ in Eq. (\ref{autocorrelacion2}), the invariance of
the trace under cyclic permutations ensures that $tr[\hat{S}_{1}^{z}(t)\hat{S%
}_{1}^{z}(0)]=tr[\hat{S}_{1}^{-}(0)\hat{S}_{1}^{z}(t)\hat{S}_{1}^{+}(0)]-%
\frac{1}{2}tr[\hat{S}_{1}^{z}(t)]$. Since $tr[\hat{S}_{1}^{z}(t)]=tr[\hat{S}%
_{1}^{z}(0)]=0$, then:%
\begin{widetext}
\begin{eqnarray}
M_{1,1}(t) &=&2\sum_{i}\frac{1}{2^{N-1}}\left\langle i\right\vert \hat{S}%
_{1}^{-}(0)\hat{U}_{LE}^{\dag }(t)\hat{S}_{1}^{z}(0)\hat{U}_{LE}^{{}}(t)\hat{%
S}_{1}^{+}(0)\left\vert i\right\rangle  \notag \\
&=&2\sum_{i\in \mathcal{A}}\frac{1}{2^{N-1}}\left\langle i\right\vert \hat{U}%
_{LE}^{\dag }(t)\hat{S}_{1}^{z}\hat{U}_{LE}^{{}}(t)\left\vert i\right\rangle
.  \label{M11_promedioEnsamble}
\end{eqnarray}%
\end{widetext}

Here, the states $\left\vert i\right\rangle $ correspond to the
computational Ising basis for $N$ spins. In the second line, we restrict the
sum to the set $\mathcal{A}$ of numbers $i$ that label basis states which
have the $1^{\mathrm{st}}$ spin pointing up, i.e. $i\in \mathcal{A}$ $%
\Leftrightarrow \hat{S}_{1}^{z}\left\vert i\right\rangle =+\frac{1}{2}%
\left\vert i\right\rangle $. Equation (\ref{M11_promedioEnsamble}) is indeed
an explicit way to rewrite Eq. (\ref{autocorrelacion1}) in the form of an
ensemble average. From the computational point of view, we are left with two
possible ways to evaluate, \textit{without any truncation,}\ the correlation
function $M_{1,1}(t)$.

The first (and naive) alternative would be the storage and manipulation of
the complete density matrix (whose size scales as $\sim 2^{N}\times 2^{N}$).
Additionally, the time-dependence represented in Eq. (\ref{autocorrelacion11}%
) would eventually require the diagonalization of the Hamiltonian matrix.
This strategy has strong limitations due to the memory constraints of any
hardware and thus one would hardly achieve systems larger than $N\sim 12$.
The second alternative would be the independent (trivially \textit{parallel}%
) computation of each of the $2^{N-1}$ expectation values in Eq. (\ref%
{M11_promedioEnsamble}). In such a case, one would handle single vectors
(whose size scales as $\sim 2^{N}$). The evolution operators, in turn, can
be implemented according to the Trotter-Suzuki formula up to a desired
accuracy \cite{Dente2013CPC}.

Remarkably, there is a \textit{parallel} way to successfully approximate the
previous calculation using quantum superpositions. Since $\hat{S}_{1}^{z}$
is a local (\textquotedblleft one-body\textquotedblright ) operator, its
evaluation in Eq. (\ref{M11_promedioEnsamble}) can be replaced by the
expectation value in a single superposition state \cite{Alv-parallelism},%
\begin{equation}
M_{1,1}(t)=2\left\langle \Psi _{neq}\right\vert \hat{U}_{LE}^{\dag }(t)\hat{S%
}_{1}^{z}\hat{U}_{LE}^{{}}(t)\left\vert \Psi _{neq}\right\rangle ,
\label{autocorrelacion7}
\end{equation}%
where: 
\begin{equation}
\left\vert \Psi _{neq}\right\rangle =\sum\limits_{k\in \mathcal{A}}\frac{%
\exp [-\mathrm{i}\phi _{k}]}{\sqrt{2^{N-1}}}\text{\ }\left\vert
k\right\rangle .  \label{neqsup}
\end{equation}%
Here, $\left\{ \phi _{k}^{{}}\right\} $ are random phases uniformly
distributed in $[0,2\pi )$. As a matter of fact, the state defined in Eq. (%
\ref{neqsup}) is a random superposition that can successfully mimic the
dynamics of ensemble calculations and provides a quadratic speedup of
computational efforts \cite{Alv-parallelism} (for similar implementations,
see also \cite{Fine2013,Pineda2014}).

\subsection{From local to global LE}

According to the basis states introduced above, the initial state in Eq. (%
\ref{rho_inicial}), can be written as%
\begin{equation}
\hat{\rho}_{0}=\sum_{j\in \mathcal{A}}2^{-(N-1)}\left\vert j\right\rangle
\left\langle j\right\vert .  \label{rho_inicial2}
\end{equation}%
Using Eqs. (\ref{rho_inicial}), (\ref{M11_promedioEnsamble}) and\ (\ref%
{rho_inicial2}), we can rewrite $M_{1,1}(t)$ as introduced in Ref. \cite%
{mesoECO-PRL1995},%
\begin{equation}
M_{1,1}(t)=2\left[ \sum_{i\in \mathcal{A}}\sum_{j\in \mathcal{A}}\frac{1}{%
2^{N-1}}\left\vert \left\langle j\right\vert \hat{U}_{LE}^{{}}(t)\left\vert
i\right\rangle \right\vert ^{2}-\frac{1}{2}\right] .  \label{autoco2}
\end{equation}%
After some manipulation,%
\begin{widetext}
\begin{equation}
\begin{split}
M_{1,1}(t)& =2\left[ \sum_{i\in \mathcal{A}}\sum_{j\in \mathcal{A}}\frac{1}{%
2^{N-1}}\left\vert \left\langle j\right\vert \hat{U}_{LE}^{{}}(t)\left\vert
i\right\rangle \right\vert ^{2}-\frac{1}{2}\right] \\
& =\sum_{i\in \mathcal{A}}\frac{1}{2^{N-1}}\left[ \left\vert \left\langle
i\right\vert \hat{U}_{LE}^{{}}(t)\left\vert i\right\rangle \right\vert
^{2}+\sum_{j\in \mathcal{A}\text{ (}j\neq i)}\left\vert \left\langle
j\right\vert \hat{U}_{LE}^{{}}(t)\left\vert i\right\rangle \right\vert
^{2}-\sum_{j\in \mathcal{B}}\left\vert \left\langle j\right\vert \hat{U}%
_{LE}^{{}}(t)\left\vert i\right\rangle \right\vert ^{2}\right] .
\end{split}
\label{contribuciones}
\end{equation}%
\end{widetext}Here, $\mathcal{B}$ stands for the complement of $\mathcal{A}$%
, i.e. $j\in \mathcal{B}$ $\Leftrightarrow \hat{S}_{1}^{z}\left\vert
j\right\rangle =-\frac{1}{2}\left\vert j\right\rangle $. We follow Ref. \cite%
{Zangara2016} to split the contributions in $M_{1,1}(t)$. Then, the first
sum in Eq. (\ref{contribuciones}) is defined as the many-body or global LE,
denoted by $M_{MB}(t)$,%
\begin{equation}
M_{MB}(t)=\sum_{i\in \mathcal{A}}\frac{1}{2^{N-1}}\left\vert \left\langle
i\right\vert \hat{U}_{LE}^{{}}(t)\left\vert i\right\rangle \right\vert ^{2},
\label{autocorrelacion51}
\end{equation}%
and stands for the average probability of revival of the many-body states.
As a matter of fact, this magnitude resembles the original LE definition
stated in Eq. (\ref{eco_jalpa}). Moreover, a widely employed extension of
the LE in many-body systems corresponds to single overlaps of specific
many-body wave functions, i.e. a specific single term in the sum of Eq. (\ref%
{autocorrelacion51}). Such an approach has been performed in many scenarios,
such as criticality \cite{Zanardi2006}, non-Markovianity in open systems 
\cite{Maniscalco2012,Wisniacki2012NM}, orthogonality catastrophe \cite%
{PollmanPRL2013}, equilibration dynamics that follows a quantum quench \cite%
{Santos2014a,Santos2014b,Santos2014c,LeaScripta2015}, many-body localization 
\cite{Santos2015prb}, among others.

The second sum in Eq. (\ref{contribuciones}) represents the average
probability of changing the configuration of any spin except the $1^{\mathrm{%
st}}$. The third sum stands for the average probability that the $1^{\mathrm{%
st}}$ spin has actually flipped, i.e. of all those processes that do not
contribute to $M_{1,1}(t)$. The sum of these terms defines a correlation
function $M_{X}(t)$,%
\begin{widetext}
\begin{equation}
M_{X}(t)=\sum_{i\in \mathcal{A}}\frac{1}{2^{N-1}}\left( \sum_{j\in \mathcal{A%
}\text{ (}j\neq i)}\left\vert \left\langle j\right\vert \hat{U}%
_{LE}^{{}}(t)\left\vert i\right\rangle \right\vert ^{2}-\sum_{j\in \mathcal{B%
}}\left\vert \left\langle j\right\vert \hat{U}_{LE}^{{}}(t)\left\vert
i\right\rangle \right\vert ^{2}\right) .  \label{autocorrelacion52}
\end{equation}%
\end{widetext}This balance of probabilities leads to the appropriate
asymptotic behavior of $M_{1,1}(t)$ according to the symmetries that
constrain the evolution. The decomposition%
\begin{equation}
M_{1,1}(t)=M_{MB}(t)+M_{X}(t)  \label{LE_contributions}
\end{equation}%
has been recently studied in Ref. \cite{Zangara2016}. The very short-time
perturbative behavior of each of these quantities has been specifically
quantified: $M_{MB}(t)$ decreases as $1-(N/4)\left( t/\tau _{\Sigma }\right)
^{2}$ and $M_{X}(t)$ increases initially as $\left( N/4-1\right) \left(
t/\tau _{\Sigma }\right) ^{2}$.  In this specific context, $\tau _{\Sigma }$
is defined as the characteristic \textit{local} time-scale of the
perturbation. The precise balance between $M_{MB}(t)$ and $M_{X}(t)$ provide
for the initial decay of $M_{1,1}(t),$ which at very short times is given by 
$1-\left( t/\tau _{\Sigma }\right) ^{2}$. These expansions indicate an
extensivity relation between $M_{MB}(t)$ and $M_{1,1}(t)$ based on the fact
that the former decays $N$ times faster than the latter. This has been
interpreted \cite{Zangara2016} as a consequence of statistical independence
(at least valid for short-times):\ the probability of refocusing a complete
many-spin state is essentially $N$ times the probability of refocusing a
single spin configuration (up or down). 

For intermediate times, the experimental evidence indicates that many-body
interactions become crucial to provide for decay of $M_{1,1}(t)$. More
precisely, reversible interactions in $\hat{H}_{0}$ are responsible for the
observed $M_{1,1}(t)$ decay rates. This experimental observation motivates
an analisys of the perturbation series beyond the short-time regime. As
already hinted in \cite{Zangara2016}, a general term in these expansions
would be proportional to $\left( t/\tau _{\Sigma }\right) ^{2}\left(
t/T_{2}\right) ^{2}$, with coefficients that increase rapidly as a function
of $N$. These terms account for the appearce of high order many-body
correlations. In the case of $M_{X}(t)$, it would indicate an increase even
faster than the one indicated above. This is fully consistent with the idea
of chaos-induced scrambling of quantum information as recently discussed in
Ref. \cite{kurchan2016quantum}. Nevertheless, this growth could not persist
indefinitely since $M_{X}(t)$ should ultimately decay, much as in a Multiple
Quantum Coherence experiment where two-spin coherences (\textit{i.e.}
correlations) should give place to four-spin coherences and so on\textbf{\ }%
\cite{munowitz1988coherence,Claudia_PhilTrans}\textbf{.}

\section{Dynamically Prepared Loschmidt echo \label{Sec_DPLE}}

\subsection{Operational ideas}

\begin{figure*}[!ht]
\centering\includegraphics[width=0.95\textwidth]{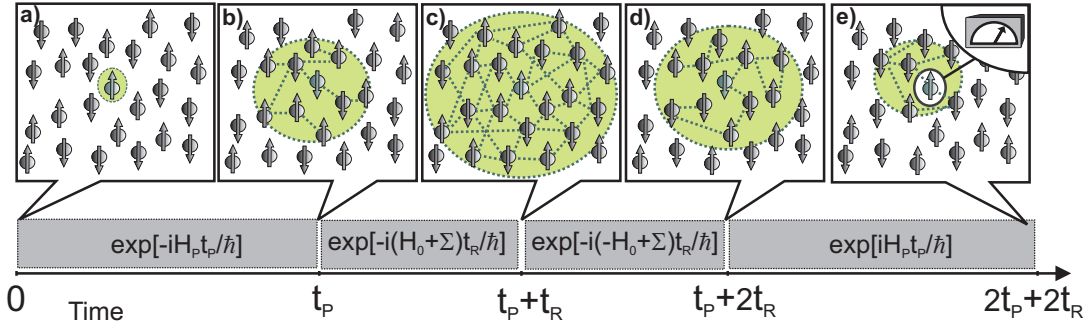} 
\caption{The preparation scheme is
included in the protocol shown in Fig. \protect\ref{Fig_esquemaLE}. (\textbf{%
a}) The initial state given by Eq. (\protect\ref{rho_inicial}) evolves
leading to a dynamically correlated state represented in (\textbf{b}). The
time reversal procedure is performed afterwards (\textbf{c})-(\textbf{d}). A
\textquotedblleft perfect\textquotedblright\ reversal of the preparation
finally leads to a local measurement in (\textbf{e}).}
\label{Fig_DPLEscheme}
\end{figure*}

Let us now discuss how to explicitly transform the local LE, i.e. $%
M_{1,1}(t) $, into a global measure of the reversibility of the many-body
dynamics. Here, the strategy is based on dynamical preparation, which
provides for an increasing complexity of the initial state in a controllable
way. Basically, the procedure consists of a sequence, schematized in Fig. %
\ref{Fig_DPLEscheme}, which is given by: preparation, standard LE, perfect
reversal of the preparation and, finally, a local measurement. The
preparation results from an evolution, controlled by a Hamiltonian $\hat{H}%
_{p}$, that occurs during a time $t_{p}$. Then, a standard LE procedure is
performed as discussed above, i.e. an imperfect forward-backward evolution.
After that, a symmetric $t_{p}$ backward evolution ruled by $-\hat{H}_{p}$
unravels the preparation, leading to a local observation. Such a local LE is
now denoted by $M_{1,1}(t,t_{p})$. The main idea here is that the local
measurement, i.e. Fig. \ref{Fig_DPLEscheme}(e), would be equivalent to the
overlap between two \textquotedblleft \textit{equilibrated}%
\textquotedblright \textit{\ }many-body wave functions that correspond to
the states of the system after preparation, i.e. the overlap of the states
in Fig. \ref{Fig_DPLEscheme}(b) and Fig. \ref{Fig_DPLEscheme}(d). In this
sense, the local autocorrelation $M_{1,1}(t,t_{p})$ would take the form of
the standard LE definition as in Eq. (\ref{eco_jalpa}), but now, in terms of
many-body wave functions.


The preparation scheme encodes the local excitation into an extended state.
Since the LE procedure is performed after such preparation, the sensitivity
under perturbations is evaluated in an initially correlated state. In order
to quantify such an observation, let us first consider a single basis state $%
\left\vert j\right\rangle $, $j\in \mathcal{A}$, that is prepared according
to $\hat{U}_{p}^{{}}(t_{p})=\exp [-\frac{\mathrm{i}}{\hbar }\hat{H}%
_{p}t_{p}] $:%
\begin{equation}
\hat{U}_{p}^{{}}(t_{p})\left\vert j\right\rangle
=\sum\limits_{k}c_{k}^{j}(t_{p})\left\vert k\right\rangle =\left\vert \Psi
_{{}}^{(j)}\right\rangle .  \label{chplg_basisstateprep}
\end{equation}%
If $\hat{H}_{p}$ does not exhibit any particular symmetry, dynamics would
not be restricted and one should assume that the summation in $k$
effectively runs over the complete Hilbert space. Moreover, the quantum
superposition in Eq. (\ref{chplg_basisstateprep}) is often assumed to be a
sort of \textquotedblleft chaotic\textquotedblright\ superposition of the
basis states $\left\vert k\right\rangle $ \cite{Flambaum1997b,Flambaum2001d}%
. In our case, this observation is particularly true if $t_{p}$ is larger
than the time needed to drive the polarization into an equilibrated value
(say, $\tau _{eq}$) \cite{Zangara2015PRA}. In such an extreme case, the
coefficients $c_{k}^{i}(t_{p})$ fluctuate so randomly as function of the
indices $j$ and $k$ that we end up replacing%
\begin{equation}
\begin{split}
\left\vert \Psi _{{}}^{(i)}\right\rangle
&=\sum\limits_{k}c_{k}^{j}(t_{p})\left\vert k\right\rangle \longrightarrow \\
& \qquad \longrightarrow \sum\limits_{k}\frac{\exp [-\mathrm{i}\phi _{k}]}{%
\sqrt{2^{N}}}\left\vert k\right\rangle =\left\vert \Phi \right\rangle ,
\end{split}
\label{reemplazo}
\end{equation}

where, as before, $\left\{ \phi _{k}\right\} $ are random phases uniformly
distributed in the interval $[0,2\pi )$. Two different realizations of these
random phases, say $\left\{ \phi _{k}\right\} $ and $\left\{ \tilde{\phi}%
_{k}\right\} $, lead to two different states,%
\begin{eqnarray*}
\left\vert \Phi \right\rangle &=&\sum\limits_{k}\frac{\exp [-\mathrm{i}\phi
_{k}]}{\sqrt{2^{N}}}\left\vert k\right\rangle , \\
\left\vert \tilde{\Phi}\right\rangle &=&\sum\limits_{k}\frac{\exp [-\mathrm{i%
}\tilde{\phi}_{k}]}{\sqrt{2^{N}}}\left\vert k\right\rangle ,
\end{eqnarray*}%
which are typically almost orthogonal:%
\begin{equation}
\left\vert \left\langle \tilde{\Phi}\mid \Phi \right\rangle \right\vert
^{2}=\left\vert \sum\limits_{k}\frac{\exp [-\mathrm{i}(\phi _{k}-\tilde{\phi}%
_{k})]}{2^{N}}\right\vert ^{2}\sim O(2^{-N}).  \label{ortogonalidad}
\end{equation}
If we replace $\hat{U}_{LE}^{{}}(t)$ by $\hat{U}_{p}^{\dag }(t_{p})\hat{U}%
_{LE}^{{}}(t)\hat{U}_{p}^{{}}(t_{p})$ in Eq. (\ref{autocorrelacion52}),%
\begin{widetext}
\begin{eqnarray}
M_{X}(t,t_{p}) &=&\sum_{i\in \mathcal{A}}\frac{1}{2^{N-1}}\left( \sum_{j\in 
\mathcal{A}\text{ (}j\neq i)}\left\vert \left\langle j\right\vert \hat{U}%
_{p}^{\dag }(t_{p})\hat{U}_{LE}^{{}}(t)\hat{U}_{p}^{{}}(t_{p})\left\vert
i\right\rangle \right\vert ^{2}-\sum_{j\in \mathcal{B}}\left\vert
\left\langle j\right\vert \hat{U}_{p}^{\dag }(t_{p})\hat{U}_{LE}^{{}}(t)\hat{%
U}_{p}^{{}}(t_{p})\left\vert i\right\rangle \right\vert ^{2}\right)  \notag
\\
&=&\sum_{i\in \mathcal{A}}\frac{1}{2^{N-1}}\left( \sum_{j\in \mathcal{A}%
\text{ (}j\neq i)}\left\vert \left\langle \Psi _{{}}^{(j)}\right\vert \hat{U}%
_{LE}^{{}}(t)\left\vert \Psi _{{}}^{(i)}\right\rangle \right\vert
^{2}-\sum_{j\in \mathcal{B}}\left\vert \left\langle \Psi
_{{}}^{(j)}\right\vert \hat{U}_{LE}^{{}}(t)\left\vert \Psi
_{{}}^{(i)}\right\rangle \right\vert ^{2}\right) .  \label{chplg_mxprep}
\end{eqnarray}%
\end{widetext}
Here comes our first specific assumption. Let us replace each of the
coherent superpositions states, namely $\left\vert \Psi
_{{}}^{(i)}\right\rangle $ and $\left\vert \Psi _{{}}^{(j)}\right\rangle $,
by incoherent superpositions as in Eq. (\ref{reemplazo}). Then, the two
summations in Eq. (\ref{chplg_mxprep}) would essentially yield the same
outcomes and this leads us to expect that $M_{X}(t,t_{p})\sim O(2^{-N})$. In
addition,%
\begin{equation}
\begin{split}
&M_{MB}(t,t_{p}) = \\
&= \sum_{i\in \mathcal{A}}\frac{1}{2^{N-1}}\left\vert \left\langle
i\right\vert \hat{U}_{p}^{\dag }(t_{p})\hat{U}_{LE}^{{}}(t)\hat{U}%
_{p}^{{}}(t_{p})\left\vert i\right\rangle \right\vert ^{2}  \notag \\
&=\sum_{i\in \mathcal{A}}\frac{1}{2^{N-1}}\left\vert \left\langle \Psi
_{{}}^{(i)}\right\vert \hat{U}_{LE}^{{}}(t)\left\vert \Psi
_{{}}^{(i)}\right\rangle \right\vert ^{2}.
\end{split}
\label{chplg_mbprep}
\end{equation}%
Here, the replacement proposed in Eq. (\ref{reemplazo}) has a practical
relevance. Indeed, when replacing $\left\vert \Psi _{{}}^{(i)}\right\rangle
\longrightarrow \left\vert \Phi \right\rangle $, Eq. (\ref{chplg_mbprep})
consists in the average of $2^{N-1}$ overlaps, each of them being
mathematically equivalent. Then, it is enough to keep just one of these
overlaps, 
\begin{equation}
M_{MB}(t,t_{p})\sim \left\vert \left\langle \Phi \right\vert \hat{U}%
_{LE}^{{}}(t)\left\vert \Phi \right\rangle \right\vert ^{2}.
\label{eq_mbentangled}
\end{equation}
If our previous observation about $M_{X}(t,t_{p})$ being exponentially small
is indeed verified, Eq. (\ref{LE_contributions}) automatically implies that $%
M_{1,1}(t,t_{p})\sim M_{MB}(t,t_{p})$. Then,%
\begin{equation}
M_{1,1}(t,t_{p})\sim \left\vert \left\langle \Phi \right\vert \hat{U}%
_{LE}^{{}}(t)\left\vert \Phi \right\rangle \right\vert ^{2}.
\label{eq_locglobent}
\end{equation}

We stress here that the previous expectations, represented by Eqns. (\ref%
{eq_mbentangled}) and (\ref{eq_locglobent}), would become valid in the limit
of $t_{p}\gg \tau _{eq}$. Additionally, we do not expect that these
equalities hold for any choice of $\hat{H}_{p}$. In particular, we need that
such a Hamiltonian can create correlations involving a large extent of the
Hilbert space $\mathcal{H}$. In practice, this justifies the specific model
adopted and described in the following Section. Furthermore, the choice is
not a purely academic issue, since it is also motivated by a possible
experimental realization.

Let us briefly comment on the case in which $\hat{H}_{p}$ does exhibit a
particular symmetry, for instance, the conservation of spin projection $[%
\hat{H}_{p},\sum_{n=1}^{N}\hat{S}_{i}^{z}]=0$. Such a case naturally induces
the decomposition of $\mathcal{H}$ in terms of subspaces $\mathcal{S}%
_{m_{z}} $ with definite spin projection quantum number $m_{z}=%
\sum_{n=1}^{N}S_{n}^{z} $. Then, the superposition in Eq. (\ref%
{chplg_basisstateprep}), $\left\vert \Psi _{\lbrack \mathcal{S}%
_{m_{z}}]}^{(j)}\right\rangle $ would be restricted to the specific
projection subspace $\mathcal{S}_{m_{z}}$ according to the initial basis
state $\left\vert j\right\rangle $. As a consequence, one should replace in
Eq. (\ref{reemplazo}) the coherent superposition $\left\vert \Psi _{\lbrack 
\mathcal{S}_{m_{z}}]}^{(j)}\right\rangle $ by a random one also defined in
such a subspace $\left\vert \Phi _{\lbrack \mathcal{S}_{m_{z}}]}\right%
\rangle $. The reasoning is analogous as before but one has to average each
of the subspaces, 
\begin{widetext}
\begin{equation}
M_{1,1}(t,t_{p})\sim (1-M_{\infty })\sum_{m_{z}}D_{m_{z}}\left\vert
\left\langle \Phi _{_{\lbrack \mathcal{S}_{m_{z}}]}}\right\vert \hat{U}%
_{LE}^{{}}(t)\left\vert \Phi _{\lbrack \mathcal{S}_{m_{z}}]}\right\rangle
\right\vert ^{2}+M_{\infty }.  \label{LEprep_porSubespacios}
\end{equation}%
\end{widetext}Here, $D_{m_{z}}$ stands for the statistical weight of the
subspace $\mathcal{S}_{m_{z}}$ and $M_{\infty }$ is the corresponding
asymptotic value of the LE. If the total spin projection in the $z$
direction is conserved (i.e. $m_{z}$ is good quantum number), then one should expect
that $M_{\infty }\sim N^{-1}$. This last asymptotic behavior occurs when the
dynamics is sufficiently complex to distribute the polarization
homogeneously among the spins in the system, as reported in \cite%
{Zangara2012}.

\subsection{The many-spin model with non-secular interactions}

\begin{figure}[!h]
\centering\includegraphics[width=0.3\textwidth]{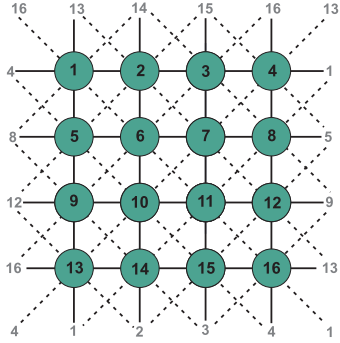} 
\caption{Square lattice with periodic
boundary conditions, $N=16$. Solid lines correspond to the spin pairs $%
\left\langle i,j\right\rangle _{\square }$ in Eq. (\protect\ref%
{eq_dipCuadrada}), and dashed lines to the pairs $\left\langle
i,j\right\rangle _{\Diamond }$ in Eq. (\protect\ref{eq_dipDiagonal}).}
\label{Fig_redCuad}
\end{figure}


In order to try out the previous expectations, we address the evaluation of
the LE in a specific spin system. As in the experiments \cite%
{patricia98,usaj-physicaA,MolPhys}, we consider a truncated dipolar
Hamiltonian,%
\begin{equation}
\hat{H}_{dip}^{\square }=\sum_{\left\langle i,j\right\rangle _{\square
}}^{N}J_{0}\left[ 2\hat{S}_{i}^{z}\hat{S}_{j}^{z}-\left( \hat{S}_{i}^{x}\hat{%
S}_{j}^{x}+\hat{S}_{i}^{y}\hat{S}_{j}^{y}\right) \right] ,
\label{eq_dipCuadrada}
\end{equation}%
where the superscript $\square $ in $\hat{H}_{dip}^{\square }$ corresponds
to the summation $\left\langle i,j\right\rangle _{\square }$ of first
nearest neighbors in a square lattice with periodic boundary conditions, as
depicted in Fig. \ref{Fig_redCuad} with solid lines. Additionally, $J_{0}$
stands for the appropriate energy units. In analogy, the dipolar coupling
between next nearest neighbors in the square lattice is given by 
\begin{equation}
\hat{H}_{dip}^{\Diamond }=\sum_{\left\langle i,j\right\rangle _{\Diamond
}}^{N}J_{0}\left[ 2\hat{S}_{i}^{z}\hat{S}_{j}^{z}-\left( \hat{S}_{i}^{x}\hat{%
S}_{j}^{x}+\hat{S}_{i}^{y}\hat{S}_{j}^{y}\right) \right] ,
\label{eq_dipDiagonal}
\end{equation}%
where, accordingly, $\left\langle i,j\right\rangle _{\Diamond }$ stands for
the pairs of spins connected by dashed lines in Fig. \ref{Fig_redCuad}.
Then, our LE procedure is defined according to the following choice: 
\begin{eqnarray}
\hat{H}_{0}^{{}} &=&\hat{H}_{dip}^{\square },  \notag \\
\hat{\Sigma} &=&\lambda \hat{H}_{dip}^{\Diamond },  \label{HamiltonianosLE}
\end{eqnarray}%
where we fix the coefficient $\lambda =0.1$.



In addition, we consider also the double-quantum (DQ) Hamiltonian \cite%
{Pines1985,Pines1987}, 
\begin{eqnarray}
\hat{H}_{dq}^{\square } &=&\sum_{\left\langle i,j\right\rangle _{\square
}}^{N}J_{0}\left( \hat{S}_{i}^{x}\hat{S}_{j}^{x}-\hat{S}_{i}^{y}\hat{S}%
_{j}^{y}\right) ,  \label{eq_dqCuadrada} \\
\hat{H}_{dq}^{_{\Diamond }} &=&\sum_{\left\langle i,j\right\rangle
_{\Diamond }}^{N}J_{0}\left( \hat{S}_{i}^{x}\hat{S}_{j}^{x}-\hat{S}_{i}^{y}%
\hat{S}_{j}^{y}\right) ,  \label{eq_dqDiagonal}
\end{eqnarray}%
with the same convention as above. This interaction, being proportional to $%
\hat{S}_{i}^{+}\hat{S}_{j}^{+}+\hat{S}_{i}^{-}\hat{S}_{j}^{-}$, does not
conserve spin projection in the $z$ direction since it mixes subspaces with $%
\delta m_{z}=2$.

Given the remarkable degree of control that can be achieved in NMR quantum
simulators, in the last years the DQ Hamiltonian has been intensively
employed to study the interplay between decoherence and correlations in
large spin arrays \cite%
{Suter2004,Boutis2012,Cappellaro2013,Claudia2014,Claudia_PhilTrans}. In
addition, it has also been employed to address localization phenomena \cite%
{Alvarez2010_localization,AlvarezScience}. Indeed, the DQ Hamiltonian can be
used to create, in a controllable way, clusters of correlated spins that can
serve as initial states for more sophisticated protocols. This motivates the
choice%
\begin{equation*}
\hat{H}_{p}^{{}}=\frac{\hat{H}_{dq}^{\square }+\hat{H}_{dq}^{_{\Diamond }}}{%
\sqrt{2}},
\end{equation*}%
which provides for the preparation dynamics.

\subsection{LE numerical evaluation}

\begin{figure}[ht]
\centering\includegraphics[width=0.5\textwidth]{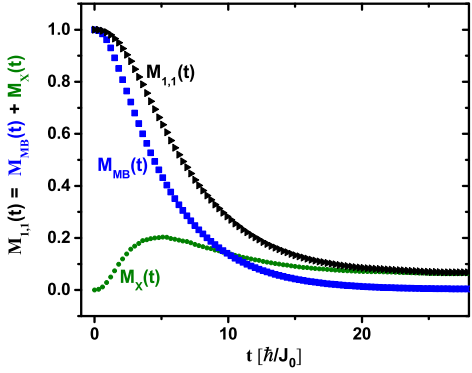} 
\caption{The local autocorrelation
function or local LE $M_{1,1}(t)$ (black triangles), and its two non-local
contributions: $M_{MB}(t)$ (blue squares) and $M_{X}(t)$ (green circles). On the one hand, $M_{MB}(t)$ exhibits a fast decay
(roughly $N$ times faster than that of $M_{1,1}(t)$) and it asymptotically goes to zero. On the other hand, $M_{X}(t)$ exhibits a 
rapid growth until it reaches a maximum, and afterwards it decays. Precisely, this decay of $M_{X}(t)$ determines the intermediate and long-time decay of $M_{1,1}(t)$ and its asymptotic value. As discussed in the text, the case considered ($N=16$) is still too small to emphasize these observations. The features in the dynamics of these correlation functions will become more prominent as $N$ increases, which may be beyond the state-of-the-art numerical techniques.}
\label{Fig_LEnum}
\end{figure}


We show in Fig. \ref{Fig_LEnum} the evaluation of the correlation functions $%
M_{1,1}(t)$, $M_{MB}(t)$ and $M_{X}(t)$, according to Eq. (\ref%
{HamiltonianosLE}). One may expect that $M_{1,1}(t)$ and $M_{MB}(t)$ differ
radically since the latter should decay, as stated above, $\sim N$ times
faster than the former. Strictly speaking, as discussed in Ref. \cite%
{Zangara2016}, $M_{1,1}^{\eta }\simeq M_{MB}$, with $\eta \simeq N/4$, which
for the case considered ($N=16$) corresponds to $\eta \simeq 4$. Such a
small exponent is basically the reason why the two correlation functions do
not separate each other considerably. An ideal finite size scaling would
involve a progression of systems satisfying $\eta \gg 1$. In short, our $%
N=16 $ is still too small. This observation indicates that, within the
state-of-the-art numerical techniques, finding numerical evidence of an
emerging PID remains a major challenge \cite{Zangara2015PRA,Zangara2016}.

\begin{figure*}[ht]
\centering\includegraphics[width=0.85\textwidth]{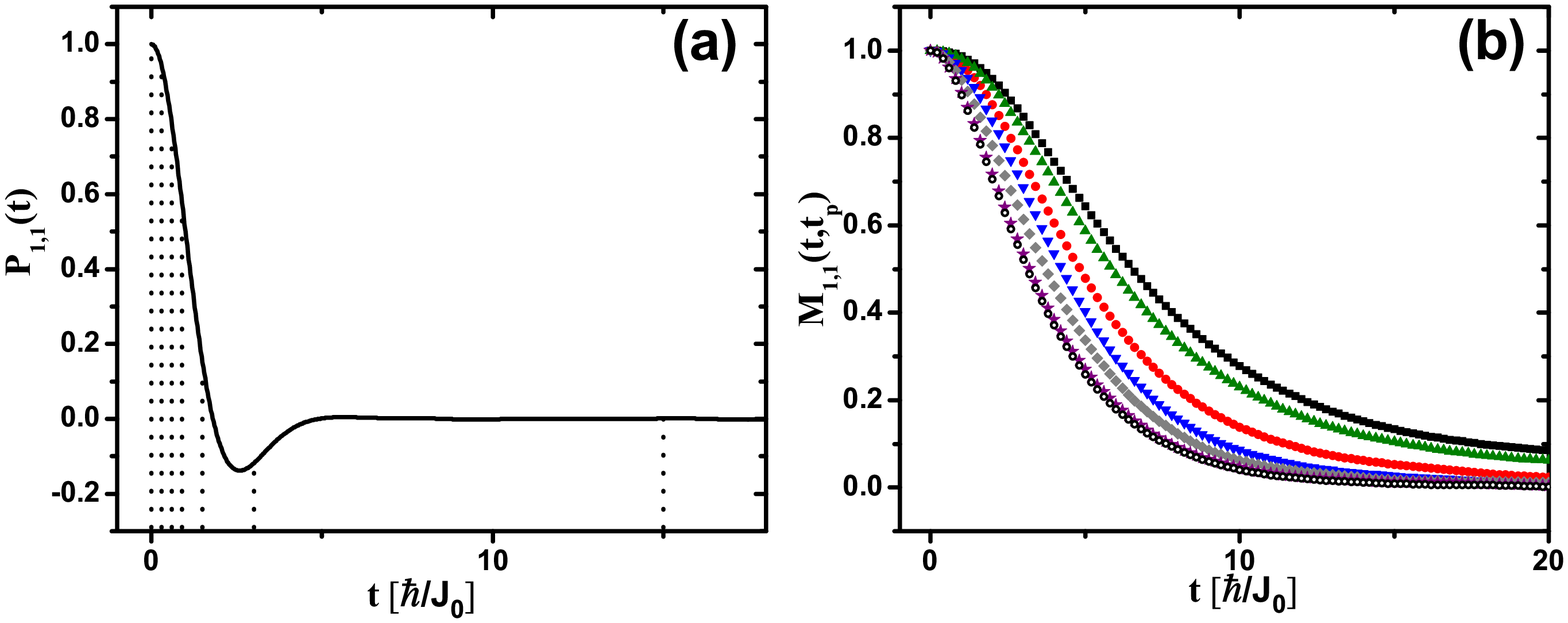} 
\caption{(\textbf{a}) Forward (local)
autocorrelation function $P_{1,1}(t)$ defined as in Eq. (\protect\ref%
{p11_prep}). Vertical dotted lines indicate the evolution times chosen to
play the role of $t_{p}$ (preparation time in the DPLE protocol). The
corresponding DPLE, $M_{1,1}(t,t_{p})$, are shown in (\textbf{b}). In
particular, $t_{p}=0\hbar /J_{0}$ (black squares); $t_{p}=0.3\hbar /J_{0}$
(green upper triangles); $t_{p}=0.6\hbar /J_{0}$ (red circles); $%
t_{p}=0.9\hbar /J_{0}$ (blue down triangles); $t_{p}=1.5\hbar /J_{0}$ (grey
diamonds); $t_{p}=3\hbar /J_{0}$ (purple stars); $t_{p}=15\hbar /J_{0}$ (open
circles). }
\label{Fig_DPLEnumeric}
\end{figure*}

We address the evaluation of the DPLE in Fig. \ref{Fig_DPLEnumeric}. In
particular, Fig. \ref{Fig_DPLEnumeric}-(\textbf{a}) shows a forward
autocorrelation function that corresponds to the dynamical preparation, i.e. 
\begin{equation}
P_{1,1}(t)=2\left\langle \Psi _{neq}\right\vert \hat{U}_{p}^{\dag }(t_{p})%
\hat{S}_{1}^{z}\hat{U}_{p}^{{}}(t_{p})\left\vert \Psi _{neq}\right\rangle .
\label{p11_prep}
\end{equation}%

The time-evolution of the local polarization at site $1$ is monitored and it
is observed that it stabilizes near zero within the characteristic time of
equilibration $\tau _{eq}\sim 5\hbar /J_{0}$. This leads us to choose
specific preparation times $t_{p}\lesssim \tau _{eq}$ and $t_{p}\gg \tau
_{eq}$. Accordingly, Fig. \ref{Fig_DPLEnumeric}-(\textbf{b}) shows $%
M_{1,1}(t,t_{p})$ for such choices of preparation times. If $t_{p}\lesssim
\tau _{eq}$, the larger the $t_{p}$, the faster the decay of $%
M_{1,1}(t,t_{p})$. When $t_{p}$ exceeds $\tau _{eq}$, a saturation regime is
observed, in which $M_{1,1}(t,t_{p})$ becomes independent of $t_{p}$. In
fact, Fig. \ref{Fig_DPLEnumeric}-(\textbf{b}) shows that $%
M_{1,1}(t,t_{p}=3\hbar /J_{0})\simeq $ $M_{1,1}(t,t_{p}=15\hbar /J_{0})$
(purple stars and open circles, respectively). In general, all the curves
representing $M_{1,1}(t,t_{p})$, for $t_{p}>\tau _{eq}$, collapse into a
single one.

The fact that $M_{1,1}(t,t_{p})$ no longer depends on the precise value of $%
t_{p}$ (provided that it exceeds $\tau _{eq}$) indicates that the specific
phases in the state $\hat{U}_{p}^{{}}(t_{p})\left\vert \Psi
_{neq}\right\rangle $ have become non-relevant for the dynamics of the
polarization. This idea can be generalized to say that outcome of a local
measurement is independent of many non-local correlations present in an
evolved many-body state. These \textquotedblleft
irrelevant\textquotedblright\ correlations are, in turn, responsible for
encoding the precise memory of the initial state, i.e. the evolution is
still unitary.


\begin{figure}[!h]
\centering\includegraphics[width=0.5\textwidth]{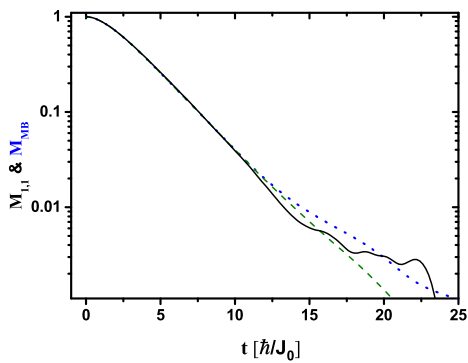} 
\caption{The DPLE. $M_{1,1}(t,t_{p}=6%
\hbar /J_{0})$, solid black line, and $M_{MB}(t,t_{p}=6\hbar /J_{0})$, blue
dotted line. The overlap of random superposition states $|\left\langle \Phi
\right\vert \hat{U}_{LE}^{{}}(t)\left\vert \Phi \right\rangle |^{2}$ is
plotted with a green dashed line. }
\label{Fig_DPLE2}
\end{figure}

In order to address the saturation aforementioned ($t_{p}>\tau _{eq}$), we
compare $M_{1,1}(t,t_{p})$ and $M_{MB}(t,t_{p})$ for $t_{p}=6\hbar /J_{0}$
in Fig. \ref{Fig_DPLE2}. In addition, as stated in Eqns. (\ref%
{eq_mbentangled}) and (\ref{eq_locglobent}), we include the overlap $%
|\left\langle \Phi \right\vert \hat{U}_{LE}^{{}}(t)\left\vert \Phi
\right\rangle |^{2}$, being $\left\vert \Phi \right\rangle $ the random
superposition state defined in Eq. (\ref{reemplazo}). The coincidence
between these three curves is remarkable. It is also worthwhile to notice
the clear exponential nature of the decay.

The previous observation essentially states the equivalence $%
M_{1,1}(t,t_{p})\simeq M_{MB}(t,t_{p})$ in the regime where $t_{p}>\tau
_{eq} $. Moreover, it identifies such saturation as the overlap between two
random superposition states that evolve ruled by perturbed and unperturbed
Hamiltonians. This is essentially the many-body extension of the
semiclassical LE definition \cite{jalpa}. Then, when the dynamical
preparation creates a sufficiently complex state, a given perturbation
yields a local LE that is same as a global one for the same perturbation.

The mentioned saturation is relevant for the picture of equilibration
discussed in Section \ref{Sec_eqloc}. When the preparation dynamics
equilibrates the polarization, the state $\hat{U}_{p}^{{}}(t_{p})\left\vert
\Psi _{neq}\right\rangle $ can be replaced by the \textquotedblleft \textit{%
equilibrated\textquotedblright } state $\left\vert \Phi \right\rangle $, as
far as the evaluation of polarization is concerned. The random phases in $%
\left\vert \Phi \right\rangle $ correspond to the fact that equilibration of
a local observable is reached when the global correlations are irrelevant or
at least redundant for such observable. A similar argument has already been
discussed precisely in the context of the LE \cite{Zangara2015PRA}, and it
provides a hint for theoretical investigations on the onset of equilibration
for local observables in closed many-body systems.

Our observations here may also provide new strategies to understand the
experimental results discussed in Sec. \ref{Sec_LEdiscussion}. Since the LE decay
accelerates as a function of $t_{p}$, the fragility of the reversal
procedure in the presence of perturbations can be systematically quantified
as a function of $t_{p}$. More precisely, this indicates that the
preparation dynamics contributes to the time-scale of the LE decay.
According to the experimental hints, this contribution would ultimately be
the dominant term in the time-scale of LE decay. Of course, this last
scenario corresponds to the TL, and thus an appropriate finite size scaling
would be needed to confirm it.


\section{Conclusions}

Starting from a conceptual and historical discussion on the microscopic
foundations of the Second Law of thermodynamics as an emergent of Classical
Mechanics, we have introduced and motivated the study of the Loschmidt echo
(LE) as tool that could help us to reveal the origin of irreversibility in a
Quantum Mechanical framework. In particular, we discussed the LE as defined
for spin systems, which corresponds to the original NMR polarization echo
experiments where the global polarization is a conserved observable. Such experimental
LE evaluations have supported an emergent, and quite paradoxical, picture
for irreversibility, here embodied by our \textit{Central Hypothesis of
Irreversibility}: in an infinite many-spin system in a highly mixed state,
any arbitrarily small perturbation is amplified by the progressively
increased complexity resulting from the many-body dynamics, which then
becomes the dominant time-scale. Thus, the reversal procedure would
ultimately be degraded within a time scale determined by such complex, but
reversible, many-body interactions.

By formulating the LE in spin systems as a local autocorrelation function,
we were able to define the non-local or global LE, and we proposed a
protocol to transform the local LE into a global one. This means that a
local LE can be employed to measure a global overlap between many-body
states. This modified LE procedure introduces a dynamical preparation of the
initial state, which creates correlations by means of an (ideally
reversible) evolution. In practice, our numerical results confirm that the
more correlated the state is, the more fragile under perturbations it
becomes, which is manifested in much shorter, perturbation dependent, time
scales. Moreover, the decay saturates at a specific time-scale that
corresponds to the global LE of random superposition states. This occurs
precisely when the preparation time $t_{p}$ exceeds the equilibration time $%
\tau _{\mathrm{eq.}}$ of the polarization. Such a saturation indicates that
the local LE no longer depends on the precise value of $t_{p}$. In other
words, a local observation of the polarization does not depend on the
specific phases encoded in the many-body evolved state.

Our observations provide for a possible way to understand the experimentally
observed PID. Indeed, we showed that reversible dynamics transforms the
original local excitation state into a more complex and sensitive one, which
in turn is shown to be much more likely to be affected by any small residual
perturbation. The time scale at which complexity is being generated could
then appear as the dominant time-scale. This dynamical regime, however, has
not yet been reached in numerical simulations, as the number of involved
spins does not seem to be large enough. Thus, while we are not yet in a
position to present a definite numerical test for the Central Hypothesis of
Irreversibility, it is worthwhile to mention that the state preparation
scheme discussed here can be implemented in a variety of actual NMR
experiments. Its application to finite and infinite systems in the form of
Loschmidt echo of the polarization- and magic echo types, could shed further
light on their strikingly different behaviors, and moreover, into the
elusive reversibility paradox.

\section{Acknowledgements}

We thank Fernando Pastawski for his careful and critical reading of the
manuscript. HMP acknowledges Alexei Kitaev for his hospitality and fruitful discussions at Caltech, and Arturo L\'opez D\'avalos and Francisco de la Cruz for early inspiring lectures on irreversibility and dissipation. We acknowledge financial support from CONICET, ANPCyT, SeCyT-UNC and MinCyT-Cor. This work used Mendieta Cluster from CCAD at UNC, that is
part of SNCAD-MinCyT, Argentina.


\end{document}